\documentstyle[12pt]{article}
\def\bcn{\begin{center}}
\def\ecn{\end{center}}
\newcommand{\beqn}{\begin{eqnarray}}
\newcommand{\eeqn}{\end{eqnarray}}
\newcommand{\epm}{\mbox{$e^+e^-$}}
\newcommand{\ee}{\mbox{$e^-e^-$}}
\newcommand{\ep}{\mbox{$e^-\gamma$}}
\newcommand{\pp}{\mbox{$\gamma\gamma$}}
\newcommand{\lc}{linear collider}
\newcommand{\sm}{standard model}
\newcommand{\br}{branching ratio}
\newcommand{\trm}{transverse momentum}

\newcommand{\cm}{center of mass}
\newcommand{\xs}{cross section}
\newcommand{\EW}{electroweak}
\def\beq{\begin{equation}}
\def\eeq{\end{equation}}
\def\barr{\begin{array}}
\def\earr{\end{array}}
\def\and{\qquad {\rm and } \qquad}
\def\etal{ {\it et al.} }
\def\ie{ {\it i.e.} }
\def\viz{ {\it viz.} }

\def\bib{\bibitem}
\def\prl#1{Phys. Rev. Lett. {\bf #1}}
\def\prd#1{Phys. Rev. {\bf D#1}}
\def\plb#1{Phys. Lett. {\bf B#1}}
\def\npb#1{Nucl. Phys. {\bf B#1}}

\def\gsw{$SU(2)_L \otimes U(1)_Y$ }
\def\wwz{$WWZ$ }
\def\wwg{$WW\gamma$ }
\def\kg{\kappa_\gamma}
\def\kZ{\kappa_Z}
\def\lg{\lambda_\gamma}
\def\lZ{\lambda_Z}
\def\gZ{g^1_Z}

\begin{document}
\thispagestyle{empty}
\setcounter{page}{0}

\begin{flushright}
MPI-PhT/94-24\\
April 1994
\end{flushright}
\bigskip
\bigskip
\bigskip
\bigskip
\bigskip

\begin{center}
{\Large\bf Electron-Electron Scattering
as a Probe of Anomalous Gauge Couplings}\\
\vspace{2em}
\large
Debajyoti Choudhury
and Frank Cuypers

{\it Max-Planck-Institut f\"ur Physik,
Werner-Heisenberg-Institut,}

{\it D--80805 M\"unchen, Germany}

\vspace{1.5ex}
{\it Emails:} {\tt debchou,cuypers@iws166.mppmu.mpg.de}
\end{center}
\bigskip
\bigskip
\bigskip

\begin{abstract}
\noindent
The effect of anomalous gauge couplings
on polarized \ee\ scattering \xs s is examined.
It turns out
that different combinations of beam polarizations
provide constraints that  are complementary to each other
and to those obtained from other experiments,
such as \epm, \ep\ and \pp\ collisions.
\end{abstract}
\newpage

\section{Introduction}

In general, electron colliders
are implicitly assumed to be
electron-positron colliders.
At one of the projected \lc s
such as CLIC, JLC, TESLA, VLEPP, {\em etc.},
this need not necessarily be the case.
Indeed,
one or both electron beams could in principle
be converted into a high energy real photon beam
by Compton scattering off an intense laser \cite{LC12}.
The new horizons opened by these possibilities
have only recently started to be explored.
But also colliding electron beams
can be obtained at no extra cost
and moreover with high degrees of polarization.
As a matter of fact,
\ee\ collisions are excellent experiments
for detecting slight deviations from the expectations
of the \EW\ sector of the \sm.
Indeed,
QCD enters the game here only at the two-loop level.
Of the many possible uses that springs to one's mind \cite{phf,wo19,hm,ccl},
not the least is an examination of the electroweak gauge sector.
This would, in principle,
be of twofold use:
first the verification of the \sm\ at the one-loop level,
and more interestingly, as a search for new physics beyond the electroweak
scale.

In a previous study \cite{wo27}
we examined the effect of anomalous
$WW\gamma$ and $WWZ$ trilinear gauge couplings
on the total \xs\ of the reaction
\beq
e^-e^- \to e^-\nu_eW^-
\label{eq1}~.
\eeq
It was seen that this novel approach gives results that are as good as
those obtainable by other methods and, more interestingly, provides a
complementary tool. The importance of the latter aspect is easily
appreciated as one attempts to unravel the potentially complex structure
of this sector of the theory.
Here we refine the aforementioned  analysis
by considering the same reaction with polarized incoming electron beams
and by taking into account the angular distribution of the outgoing electron.
The use of polarized beams
provides two complementary experiments
which turn out to explore different regions
of the five-dimensional parameter space considered here.
The resolving power of the reaction
is also significantly enhanced by the study of angular distributions
instead of just total \xs s.

We plan the rest of the article as follows. In section \ref{anomcoupl} we
give a brief introduction to the possible anomalous couplings, consistent
with certain symmetry properties, that are relevant for the interaction
(\ref{eq1}). We also indicate the constraints (direct and indirect) that
have so far been obtained in the literature. Section \ref{SMexpect}
examines the \sm\ expectations for the above process in some detail. This
indicates the relevant kinematical cuts that lead to a more efficient
extraction of the signal. In section \ref{calcul} we begin by examining
the quantitative effect that the presence of anomalous couplings have on
the total cross section. This is followed by a discussion on the angular
distributions and the statistical significance of any bounds thus obtained.
Section \ref{results} contains our results in the form of contours of
detectability for different planes of the five-dimensional parameter space.
Our conclusions are then summarized in section \ref{concl}.

\section{Anomalous Couplings}
         \label{anomcoupl}

In the \sm, the tree-level
self-couplings of the vector bosons are of
course dictated by the gauge principle. Deviations from such couplings could
arise from two different (but inter-related) sources : $(i)$ quantum
corrections and $(ii)$ the presence of a new threshold nearby the weak scale
and/or the ``gauge bosons'' themselves being composite objects. The quantum
corrections in the \sm\ have been investigated \cite{couture}
and turn out to be
rather small. Precisely due to this, any evidence of deviation from the
standard couplings would be an unmistakable signature for the existence of
a new physics realm, thus delivering us from the pessimistic `Big Desert'
scenario.

An obvious arena for detecting such effects would be to look for their
effect on precisely measured low--energy processes. Indeed such
studies\cite{derujula} had
initially claimed that precision measurements at LEP had already ruled out
deviations large enough to be directly measurable in the near future. These
analyses were criticised
though on two accounts\cite{einhorn}: $(i)$ the use of a
\gsw gauge dependent Lagrangian; and $(ii)$ an improper use of the cut-off
procedure to regulate divergent integrals. A way out of this impasse was
pointed out by Burgess and London\cite{burgess} who showed that
any Lorentz and electromagnetic  gauge invariant Lagrangian for
 $W$'s and  $Z$'s
 automatically obeys
\gsw gauge invariance, realized nonlinearly in general. Furthermore, only the
logarithmic divergences are relevant for the issues at hand. Subsequent
reexaminations, both in the case of a non-linearly realized \gsw symmetry
with an arbitrary symmetry breaking sector\cite{crs},  or that of a
decoupling theory \cite{zeppen} show that the constraints imposed by LEP data
are considerably weaker than claimed previously\footnote{In fact,
if such an analysis were to be
naively extended beyond the region of validity of a momentum expansion,
the constraints would turn out to be rather weak \protect\cite{crs}.}.
This methodology  however
has an inherent ``weakness'' in that the answers depend crucially on the
assumptions made about the higher theory. Such exercises are therefore no
substitute for direct measurements.

In our discussion, we shall restrict ourselves only to \wwz and \wwg
couplings.
Analogous studies for quartic boson couplings \cite{bb}
or for triple neutral boson vertices \cite{neutral}
have been performed by other authors.
We shall furthermore restrict
the analysis to  $C$ and $P$ conserving couplings. Though
$CP$--violating
terms in the Lagrangian would contribute to the cross sections that we are
interested in, these are best isolated by looking at final state
asymmetries\cite{biswarup}. The effective Lagrangian parametrizing the $WWV$
vertex ($V = \gamma,Z$) can then be expressed as\cite{peccei}
\beq
     {\cal L}_{\it eff}^{WWV}= -i g_{V}
             \left[ g^1_V
               \left( W^\dagger_{\alpha \beta} W^\alpha
                      - W^{\dagger\alpha} W_{\alpha \beta}
                \right) V^\beta
             +
               \kappa_V  W^\dagger_{\alpha} W_\beta
                               V^{\alpha\beta}
            + \frac{\lambda_V}{M_W^2}
                 W^\dagger_{\alpha \beta} {W^\beta}_\sigma
                 V^{\sigma\alpha} \right]
      \label{lagrangian}
\eeq
where $V_{\alpha\beta} = \partial_\alpha V_\beta - \partial_\beta
V_\alpha $ and  $W_{\alpha\beta} = \partial_\alpha W_\beta -
\partial_\beta W_\alpha $.

In Eq.~(\ref{lagrangian}), $g_V$ is   the $WWV$ coupling in
the \sm\,
\viz,
\beq
g_\gamma = e, \qquad g_Z = e \cot \theta_W
\eeq
where $\theta_W$ is the weak mixing angle. Of the other couplings, only
$g^1_\gamma$ is constrained by electromagnetic gauge invariance to be
unity, and the rest are model dependent. In the \sm\, we have
\beq
\gZ = \kg = \kZ =1  \and \lg = \lZ = 0.
\eeq
It should be noted that the couplings $\kappa_V$, $\lambda_V$ and $\gZ$
are not really constants but actually represent form factors and hence, in
principle,  are energy--dependent. It can then be seen easily that even
higher dimensional operators can essentially be written in terms of these
quantities.

Till date, the only direct constraints are those obtained  for $\kg$ and $\lg$
from an analysis of the process $p\bar{p} \rightarrow e
\nu \gamma X$ at the Collider Detector at Fermilab (CDF) \cite{CDF} and at
UA2 at CERN \cite{UA2}. Assuming these events to have arisen from $W$--decay
consequent to $W\gamma$ production, parts of the $\kg$--$\lg$ plane could
be ruled out \cite{rahul-1}. The constraints are still very weak though,
especially when compared to the one--loop contributions in the
\sm\ \cite{couture}.

The above limits can be improved somewhat though at HERA \cite{Snowmass}, but
owing to the relatively low luminosity and energy and also due to large
background noise, strong limits on the $Z$--couplings are not expected. The
situation is poised for dramatic improvement both in the context of upcoming
machines like LEP-200 and the LHC \cite{LEP2,kane} as well as future
linear colliders operating either in the traditional $e^+ e^-$
mode \cite{rahul-nita}, or as $e \gamma$ and $\gamma \gamma$
colliders \cite{photon}.

\section{Standard Model Expectations}
        \label{SMexpect}

\begin{figure}[htb]
  \vskip 10in\relax\noindent\hskip -1in\relax{\includegraphics{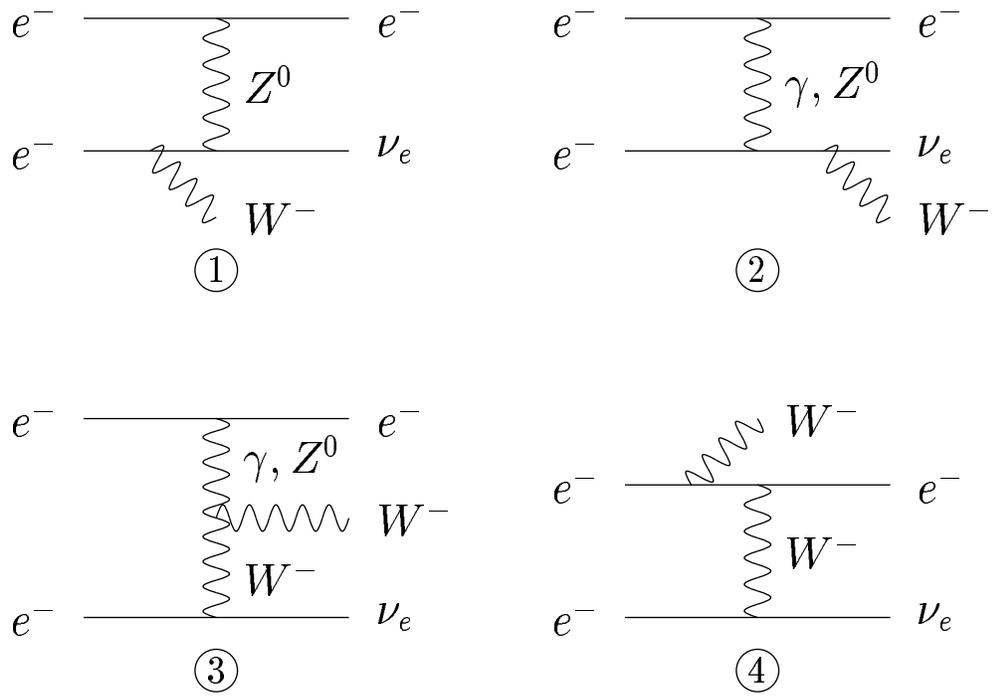}}
\vskip -4in
\caption{
Lowest order Feynman diagrams
contributing to the process (\protect\ref{eq1}).
}
\label{feyn}
\end{figure}

The Feynman diagrams contributing at lowest order to the scattering
process (\ref{eq1})
are depicted in Fig.~\ref{feyn}.
If both electron beams are right-polarized
the process does not take place,
because at least one of the fermion lines
is connected to a $W$ boson.
For the $LR$ combination of initial helicities,
diagram 4,
in which a $W$ is exchanged between the two fermion lines,
does not contribute.
For the $LL$ configuration,
the same diagrams with the initial electrons exchanged
have to be added. 
In all calculations we ignore the electron mass,
as is justified at the energies considered.

The diagrams with a virtual photon
induce a collinear singularity
which is  regulated only by the electron mass.
Nevertheless,
this dangerous region of phase space
is naturally avoided by a simple acceptance cut
on the angle of the emerging electron with the beam axis.
Throughout the rest of this paper
we have thus imposed the rapidity cut
\beq
|\eta(e^-)| < 3
\qquad{\rm or}\qquad
|\cos\theta(e^-)| < .995
\qquad{\rm or}\qquad
5.7^\circ < \theta(e^-) < 174.3^\circ
\label{eq4}~.
\eeq
To ensure that the outgoing electron is indeed visible,
we also required a minimum energy of 10 GeV:
\beq
E(e^-) > 10~{\rm GeV}
\label{eq6}~.
\eeq

There are several sources of background to the process (\ref{eq1}).
Leptonic decays of the $W^-$
cannot be distinguished from non-resonant lepton production
$e^-e^- \to e^-\nu_e \ell^-\nu_\ell$,
which proceeds via the exchange of a virtual $W^-$ and a $Z^0$ or $\gamma$.
In order to avoid this irreducible background altogether,
we thus concentrate only on the hadronic decays of the $W^-$.
In this case,
the competing non-resonant mechanisms
can be reduced to an innocuous level
by an invariant mass measurement of the jets,
the very act of $W$ reconstruction.
Still $Z^0$ Bremsstrahlung
$e^-e^- \to e^-e^-Z^0$
could be mistaken for our signal
if one of the outgoing electrons disappears along the beampipe.
To avoid this background
we have also demanded an imbalance in the observed \trm\ of the event:
\beq
p_\perp(e^-+W^-) > 10~{\rm GeV}
\label{eq5}~.
\eeq

\begin{figure}[htb]
  \vskip 10in\relax\noindent\hskip -1.5in\relax{\includegraphics{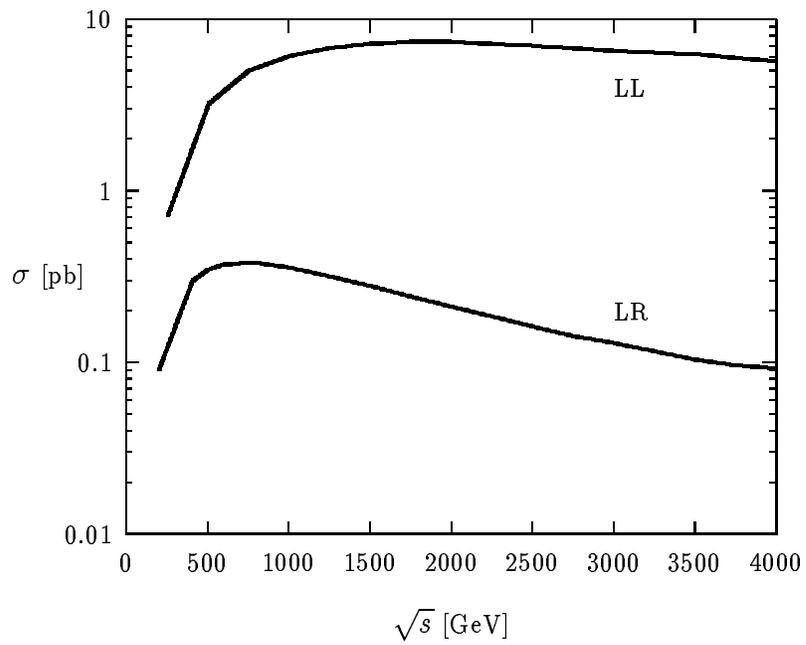}}
\vskip -4in
\caption{
Standard model cross sections of the process (\protect\ref{eq1})
as a function of the \cm\ energy
for both possible combinations of beam polarizations.}
   \label{energy dep}
\end{figure}

The dependence of the resulting \sm\ \xs\
as a function of the \cm\ energy
is shown in Fig.~\ref{energy dep}
for the two combinations of beam polarizations\footnote{Since very high
 degrees of longitudinal polarization
should be available at future linear colliders, for simplicity
we  assume 100\%\ polarization in the following.
The effects of dilution can be easily incorporated, though.},
$LL$ and $LR$.
Clearly,
the $LL$ configuration dominates,
so that almost all conclusions drawn for it
also apply as well for unpolarized beams.
Note the wide threshold region,
which is typical for vector boson production.
The decrease of the \xs\ at large energies
is due to the cut (\ref{eq4}).

It is interesting to dwell a little
on the expected topology of these events,
which is significantly different
for each combination of initial polarizations
as well as at low and high energy.

\subsection{$LR$ Polarization}
To facilitate the following discussions,
let us label for the time being
the two incoming left-handed and right-handed electrons respectively
$e_L$ and $e_R$
and the outgoing particles
$e$, $W$ and $\nu$.

\begin{figure}[htb]
\vskip 8in\relax\noindent\hskip -1.5in\relax{\includegraphics{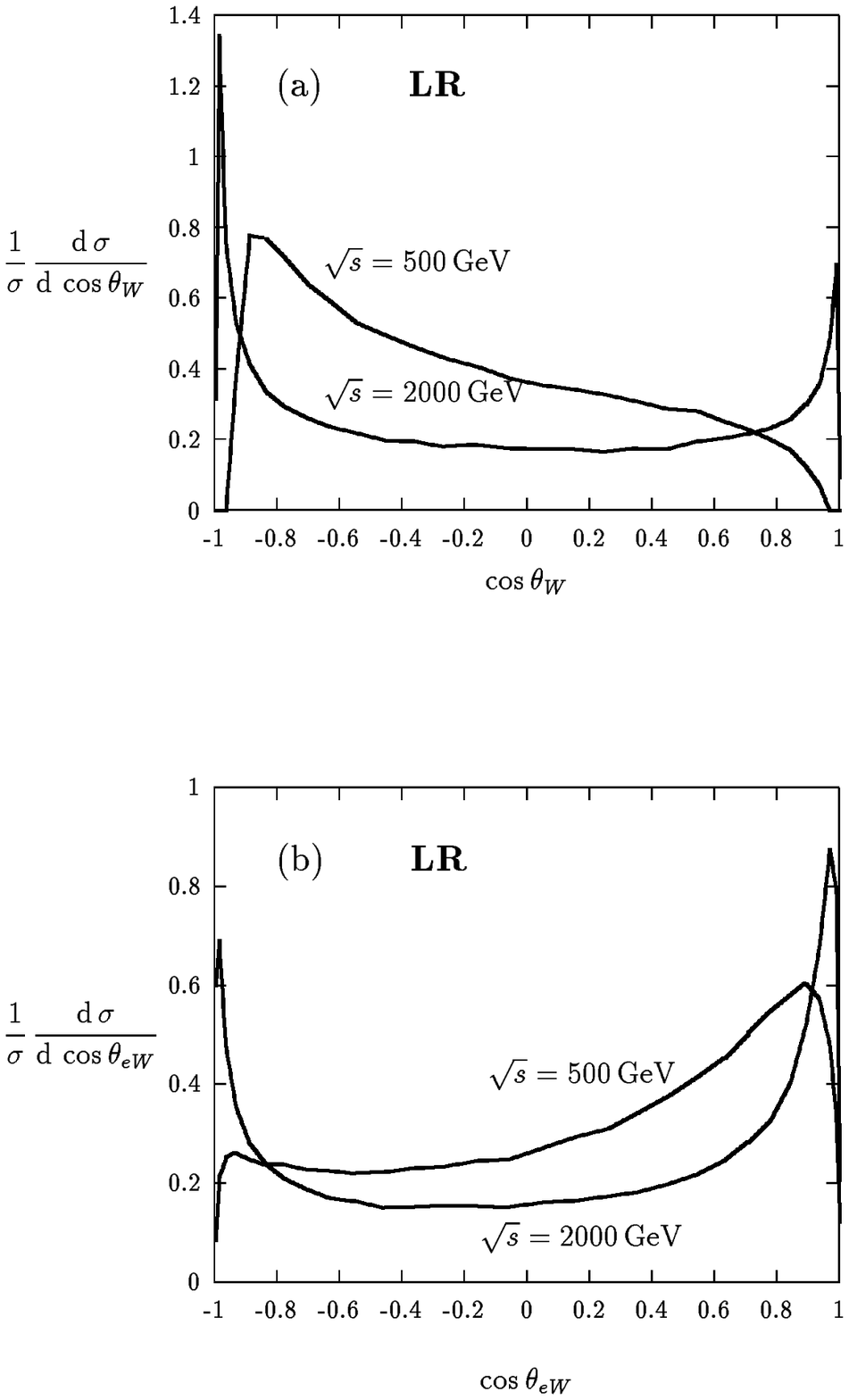}}
\caption{
Standard model low and high energy distributions of
(a) the angle of the $W^-$ with respect to the left-polarized beam,
(b) the angle spanning the emerging $W^-$ and $e^-$.
One of the beams is left-,
the other right-polarized.}
      \label{sm_lr}
\end{figure}

At low energy,
the dominant diagrams
are those which involve a photon exchange,
{\em i.e.} diagrams 2 and 3.
Because of the low \trm\ transfer of the photon,
$e$ stays close to the original trajectory
of $e_R$.
The rest of the reaction can then be seen as
an electron-photon $t$-channel scattering
$e_L \gamma^* \to \nu W$,
where the photon is almost on-shell
and aligned with $e_R$.
In the $e_L \gamma^*$ \cm\ frame,
$W$ strongly tends to emerge
opposite to $e_L$ \cite{wo17}.
This trend remains
even after boosting back to the laboratory frame,
as is shown in Fig.~\ref{sm_lr}a.
The observed $e$ and $W$
tend thus to emerge
into the detector
in the same direction,
as can inferred from Fig.~\ref{sm_lr}b.
Note, however,
the little bump in the distribution at larger angles,
which is due to the $Z^0$ exchange diagrams.

At high energy,
all diagrams contribute with comparable strength.
Still $e$ tends to be even less scattered
away from the direction of $e_R$.
However,
the first diagram of Fig.~\ref{feyn}
induces now an important fraction of events
where $W$ is emitted in the same direction as the initial $e_L$,
as can be seen in Fig.~\ref{sm_lr}a.
The observed $e$ and $W$
can thus now be emitted both parallel as well as  anti-parallel.
This is shown in Fig.~\ref{sm_lr}b,
where the low-energy bump
becomes a peak.

The energy of $e$ and especially $\nu$
is always strongly peaked close to the beam energy.
In contrast,
the energy of $W$ is distributed more broadly,
with a strong tendency to be emitted at rest,
though.

\subsection{$LL$ Polarization}
\begin{figure}[htb]
\vskip 8in\relax\noindent\hskip -1.5in\relax{\includegraphics{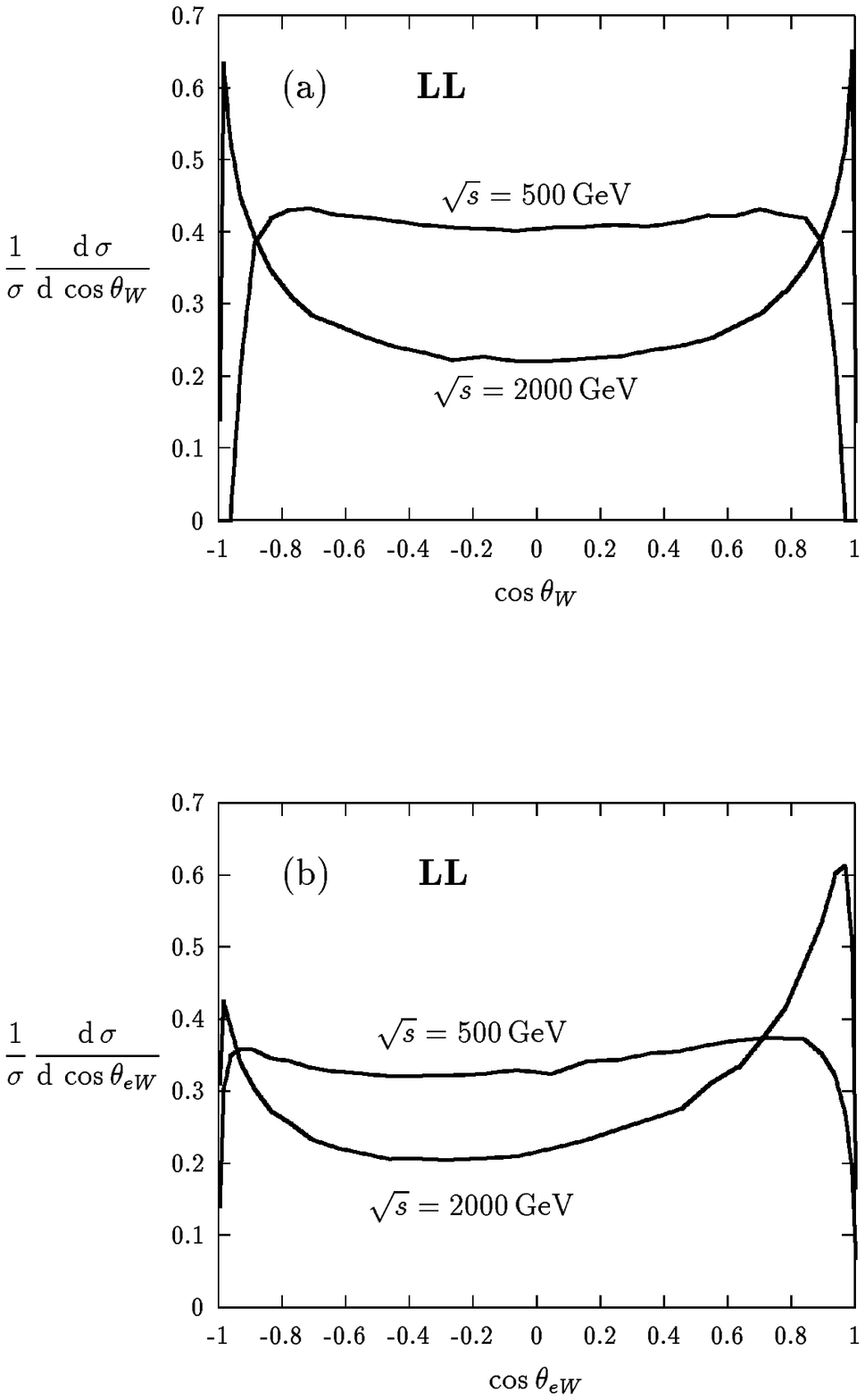}}
\caption{
Same as Fig.~\protect\ref{sm_lr},
for both beams being left-polarized.}
      \label{llwel}
\end{figure}

The picture is now tainted by the interferences
of diagram 4 of Fig.~\ref{feyn} with the three others.
At low energies,
the angular distributions in Figs~\ref{llwel}
can still be explained
by the competing trends of diagrams 2, 3
and diagram 4.
The latter favours events
with $W$ and $e$ back-to-back
and transverse to the beam direction.
In contrast,
diagrams 2 and 3,
similarly to the $LR$ case,
favour collinear topologies.
The resulting $e$ angular distribution
is peaked along the beampipe,
whereas the $W$ and $We$ angular distributions are rather flat.

At high energies,
the interferences become strongly destructive
and it stops making sense to talk of the contributions of individual diagrams.
The general trend is the same as at low energies,
just with a more enhanced peaking.

As for the $LR$ case
the energy of $e$ and especially $\nu$
is strongly peaked close to the beam energy,
whereas the energy of $W$ is distributed more broadly,
with a strong tendency to be emitted at rest.

\begin{figure}[htb]
  \vskip 10in\relax\noindent\hskip -1in\relax{\includegraphics{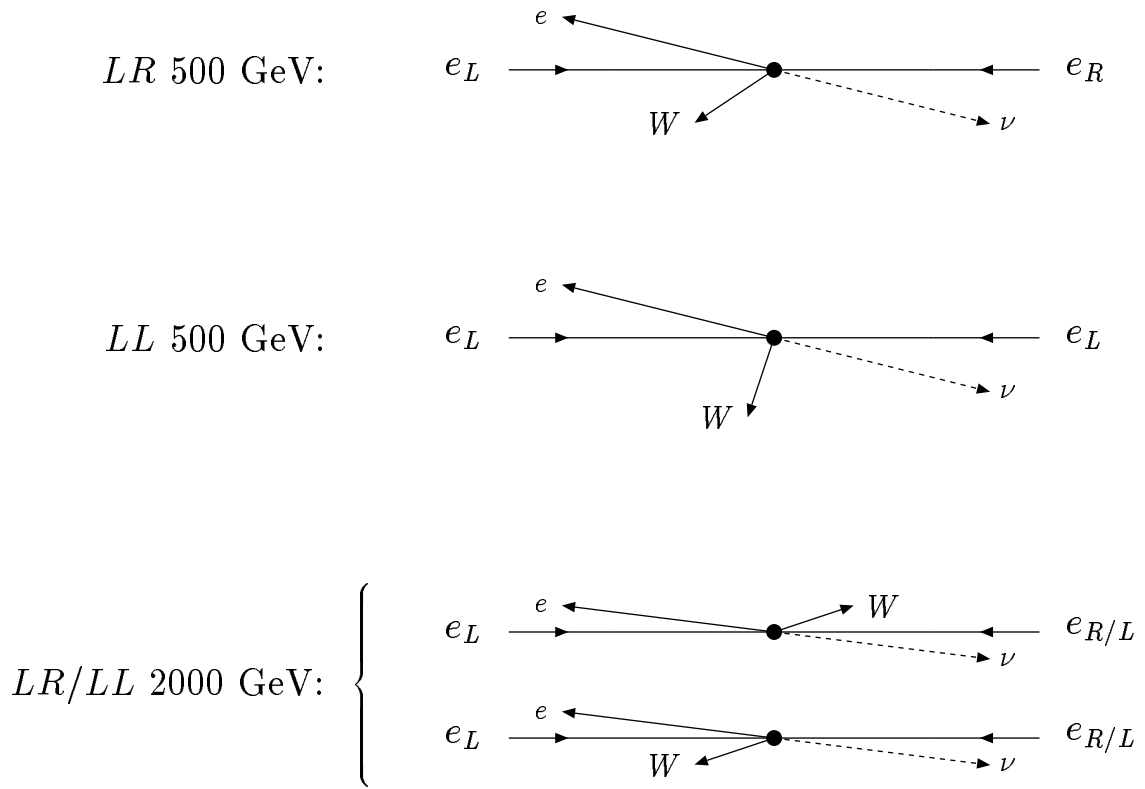}}
\vskip -4in
\caption{
Typical event topologies. The arrows represent the three--momenta of the
particles.}
\label{topo}
\end{figure}

Examples of typical event topologies
for both polarization modes
are depicted in Fig.~\ref{topo}.

\section{Calculations}
    \label{calcul}

In the following,
we have only considered a \cm\ energy of 500 GeV,
restricting our study to the case of a typical \lc\
of the next generation.
In this case
the \sm\ \xs\ is 3.19 pb for $LL$ beams
and 0.348 pb for $LR$ beams.

\begin{figure}[htb]
\vskip 8in\relax\noindent\hskip -1.5in\relax{\includegraphics{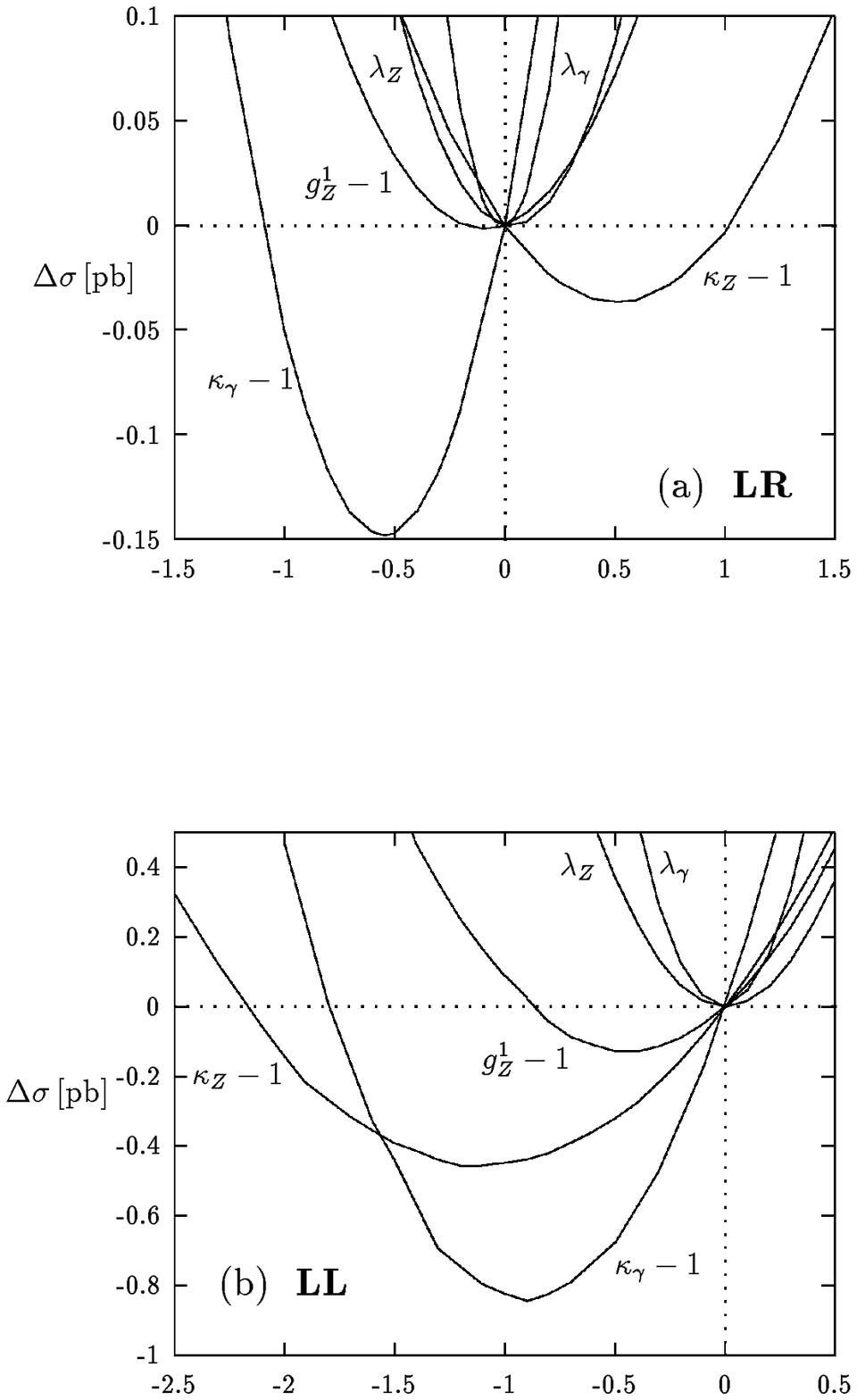}}
\caption{
Contribution to the total cross section
of each of the anomalous couplings,
while all others are held to their \sm\ values.
The \sm\ cross sections
are 3.19 pb for $LL$ polarization and 0.348 pb for $LR$.}
  \label{parabola}
\end{figure}

The dependence of the total cross sections on each of the couplings can be
seen from Figs~\ref{parabola}. For each of these curves, all the other
couplings have been held to their standard model values. As is easily seen,
the dependence on the \wwg couplings is more pronounced than on the
\wwz couplings,
irrespective of the beam polarization.
This can be traced back to the higher mass of the $Z^0$.

Note that while the \xs\ as a function of the dimension six
operators is nearly symmetric about the \sm\ value of $\lambda_{\gamma,Z} = 0$,
this certainly is not the case for the other three. Thus a measurement of
the total \xs\ alone cannot resolve potentially large values of these
couplings from the \sm\ point. This is precisely the reason why the
consideration of angular distributions is important. Before we embark on that,
note that the dependence of \xs\ on $\kZ$ and $\gZ$ is quite different for the
two beam polarizations. This will obviously be important in unravelling
the \wwz couplings from the \wwg ones.

\begin{figure}[htb]
\vskip 8in\relax\noindent\hskip -1.5in\relax{\includegraphics{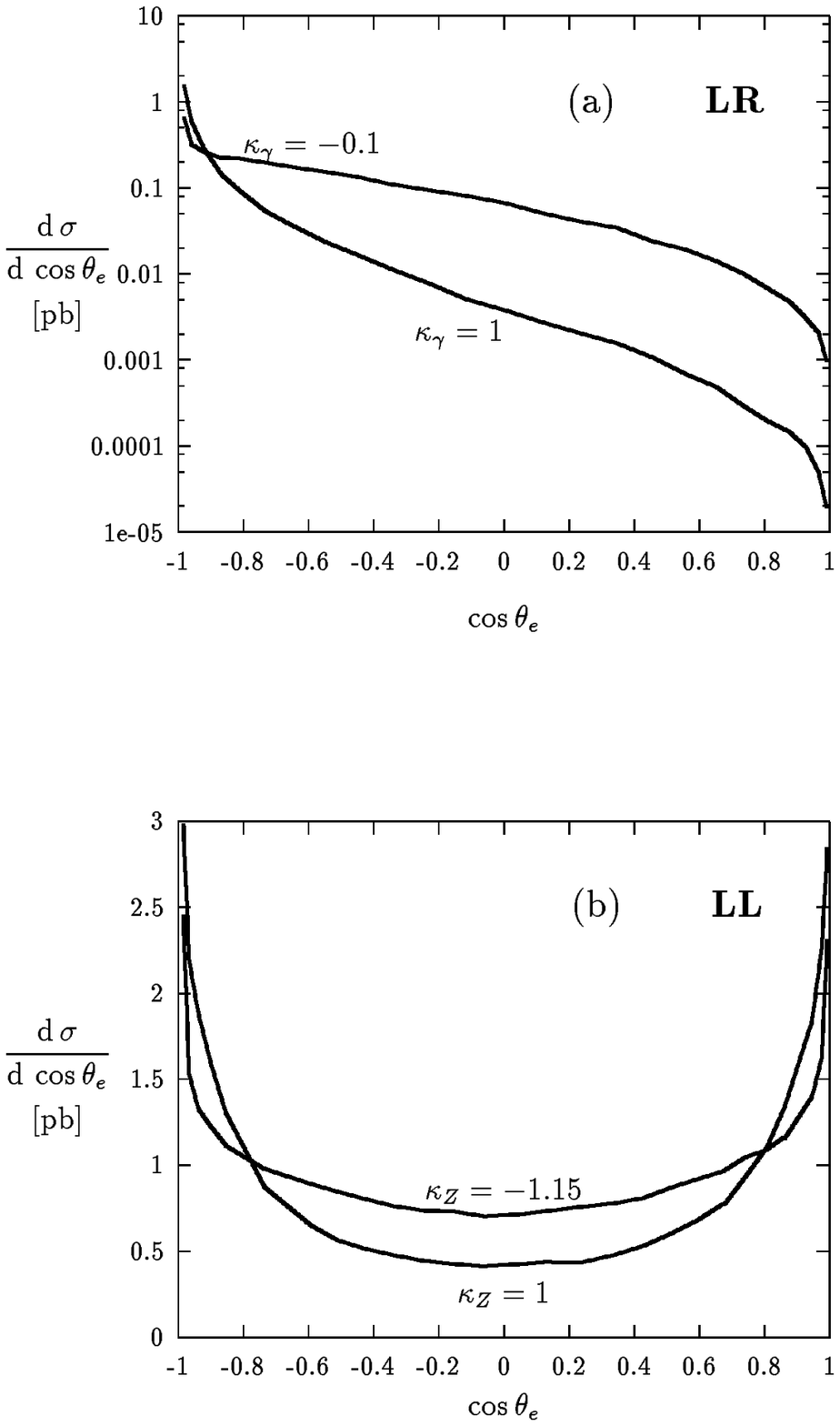}}
\caption{
Differential distributions for particular values of the anomalous
couplings
which lead to the same integrated cross section as
the one expected from the \sm. }
       \label{distrib}
\end{figure}

The \sm\ differential \xs\
as a function of the angle of the emerging electron
is displayed for both polarization schemes
in Fig.~\ref{distrib}.
The poles in the directions of the beams
are never reached,
thanks to the angular cut (\ref{eq4}).
Also shown are the  distributions
for one particular anomalous value of $\kappa_V$ each ($\kZ = -2.15 $
for $LL$ and $\kg = -1.1$ for $LR$).
Although they differ significantly from the \sm\ distributions,
the total \xs\ is almost the same in either case.
These examples show how important
it is to consider angular correlations.

In the following we shall thus
subdivide the angular range (\ref{eq4})
in 20 bins for the $LL$ case
and in 10 bins for the $LR$ case.
For a given choice of anomalous gauge parameters
we compare the content of each bin
with the \sm\ expectation.
Instead of the angle of the emerging electron
with the beam axis,
one could as well have chosen another variable,
such as the angle of the $W^-$ or the $\nu_e$,
or the energy or \trm\ of any combination of the particles.
None of these, though,
can be measured with as little systematic uncertainties\footnote{
	It turns out that the polar angle of the $W^-$ is actually
	a more sensitive variable than the polar angle of the $e^-$.
	If its resolution can be improved,
	its use might thus yield even more precise results.
}.

To estimate the resolving power of the reaction (\ref{eq1}),
we compute the $\chi^2$ function
\beq
\chi^2
=
\sum_{i=1}^{bins}
\left|
	{N_{SM}(i) - N_{anom}(i)
	\over
	\Delta N_{SM}(i)}
\right|^2
\label{eq2}~,
\eeq
where $N_{SM}$ is the number of events predicted by the \sm\
and $N_{anom}$ is the number of events effectively measured
in the presence of anomalous gauge couplings.
The sum runs over all 10 ($LR$) or 20 ($LL$) bins
in $\theta(e^-)$
over the allowed range (\ref{eq4}).
The number of events
$N = {\cal L} \sigma \epsilon_W$
is calculated assuming a typical design integrated luminosity
${\cal L} = 10$ fb$^{-1}$
and a \br\ for hadronic decays of the $W^-$ of $\epsilon_W \approx 2/3$.
The error in Eq.~(\ref{eq2})
is a combination of statistical and systematic errors
\beq
\Delta N = \sqrt{ (\sqrt{N})^2 + (\delta_{syst}N)^2 }
\label{eq3}~.
\eeq
Since the emergence angle of an electron should be measurable
to an accuracy better than 10 mrad,
the main systematic errors originate
from the luminosity measurement
and from detector efficiencies.
Since these should not exceed 1\%,
we conservatively assume an overall systematic uncertainty of 2\%
(\ie $\delta_{syst}=.02$ ).
The error is thus dominated by the statistical fluctuations
which amount to an average per bin
of 2.5\%\ for the $LL$
and 5.5\%\ for the $LR$ beam polarizations.

To avoid spurious results,
a bin is rejected from the analysis
$(i)$   if the difference between the \sm\ expectation
	and the measured number of events
	is less than one
and
$(ii)$  if the \sm\ expectation is less than one event
	while the measured number is less than three.
In order not to loose significant statistics
because of these restrictions,
we have used less angular bins for the $LR$ initial polarizations.

\section{Results}
    \label{results}

\begin{figure}[htb]
\vskip 10in\relax\noindent\hskip -1.2in\relax{\includegraphics{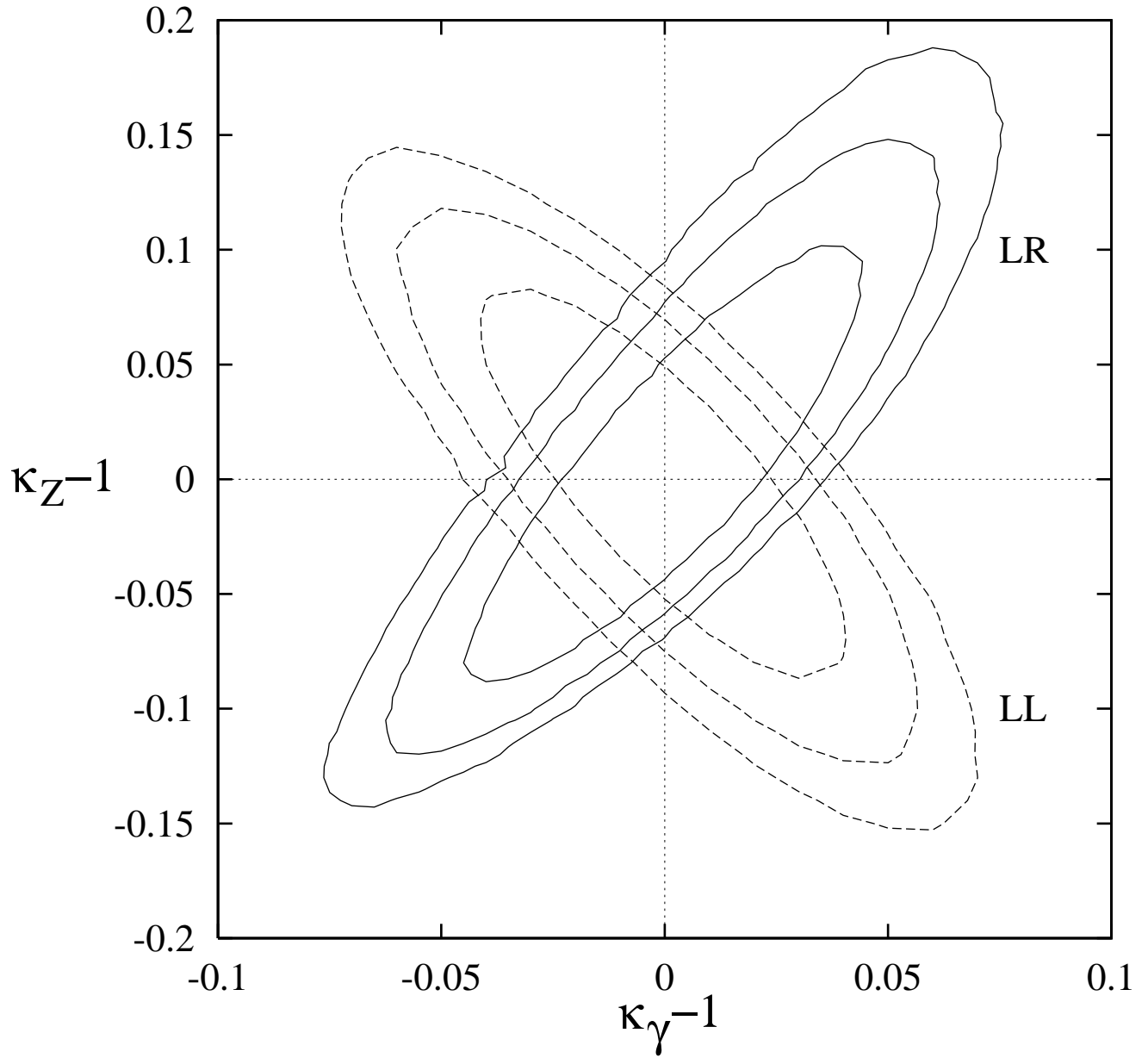}}
\vspace{-30ex}
\caption{Contours of detectability at different confidence levels
         ($\chi^2 = 2,4,6$ --- see Eq.~(\protect\ref{eq2})) in the $\kg$--$\kZ$
         plane for $LL$ (dashed lines) and $LR$ (solid lines)
         polarizations. The remaining couplings assume their \sm\ values.}
       \label{kgkz}
\end{figure}

\begin{figure}[htb]
\vskip 10in\relax\noindent\hskip -1.2in\relax{\includegraphics{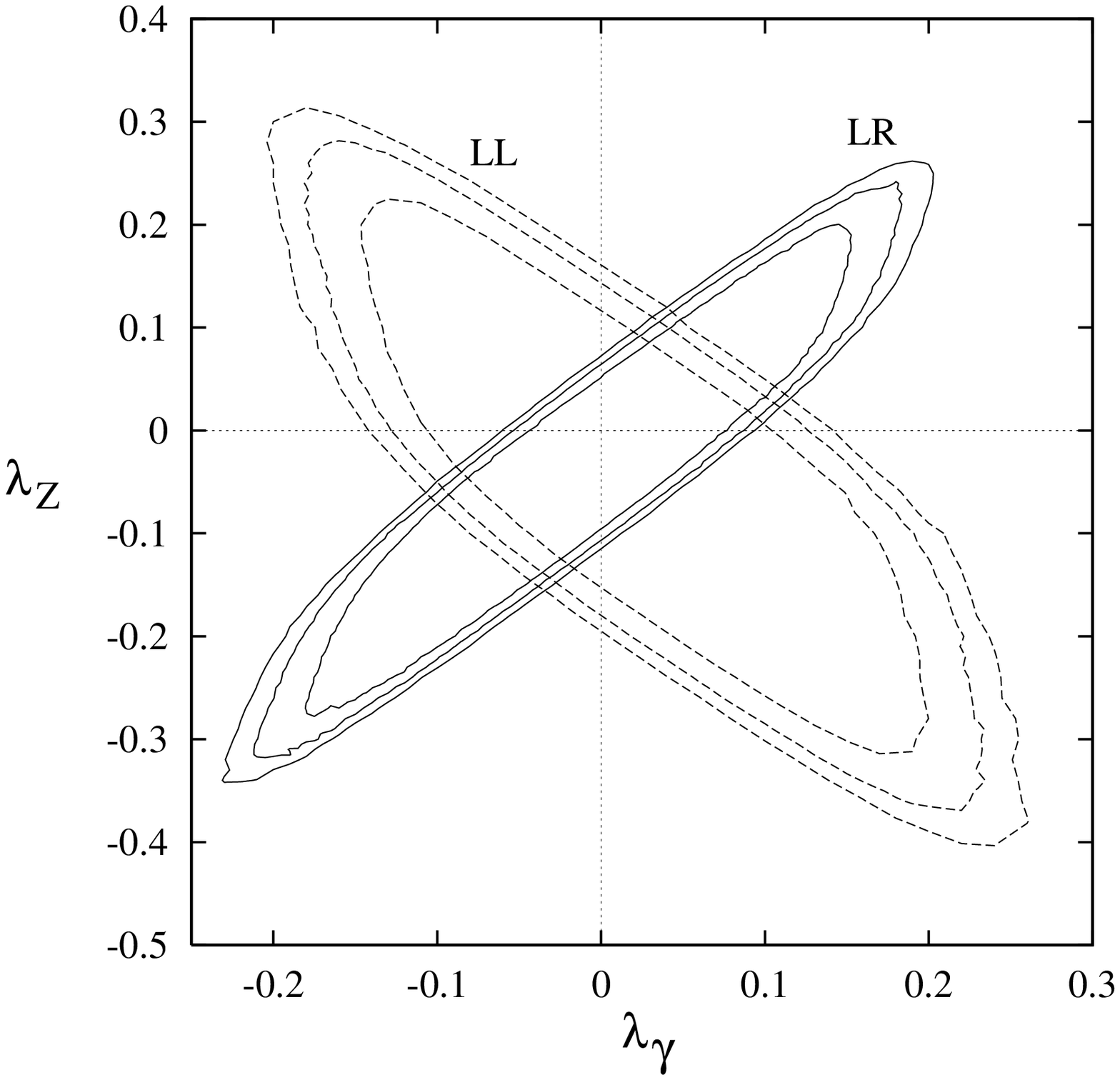}}

\vspace{-30ex}
\caption{As in Fig.~\protect\ref{kgkz}, but in  the $\lg$--$\lZ$
         plane instead.}
       \label{lglz}
\end{figure}

\begin{figure}[htb]
\vskip 10in\relax\noindent\hskip -1.2in\relax{\includegraphics{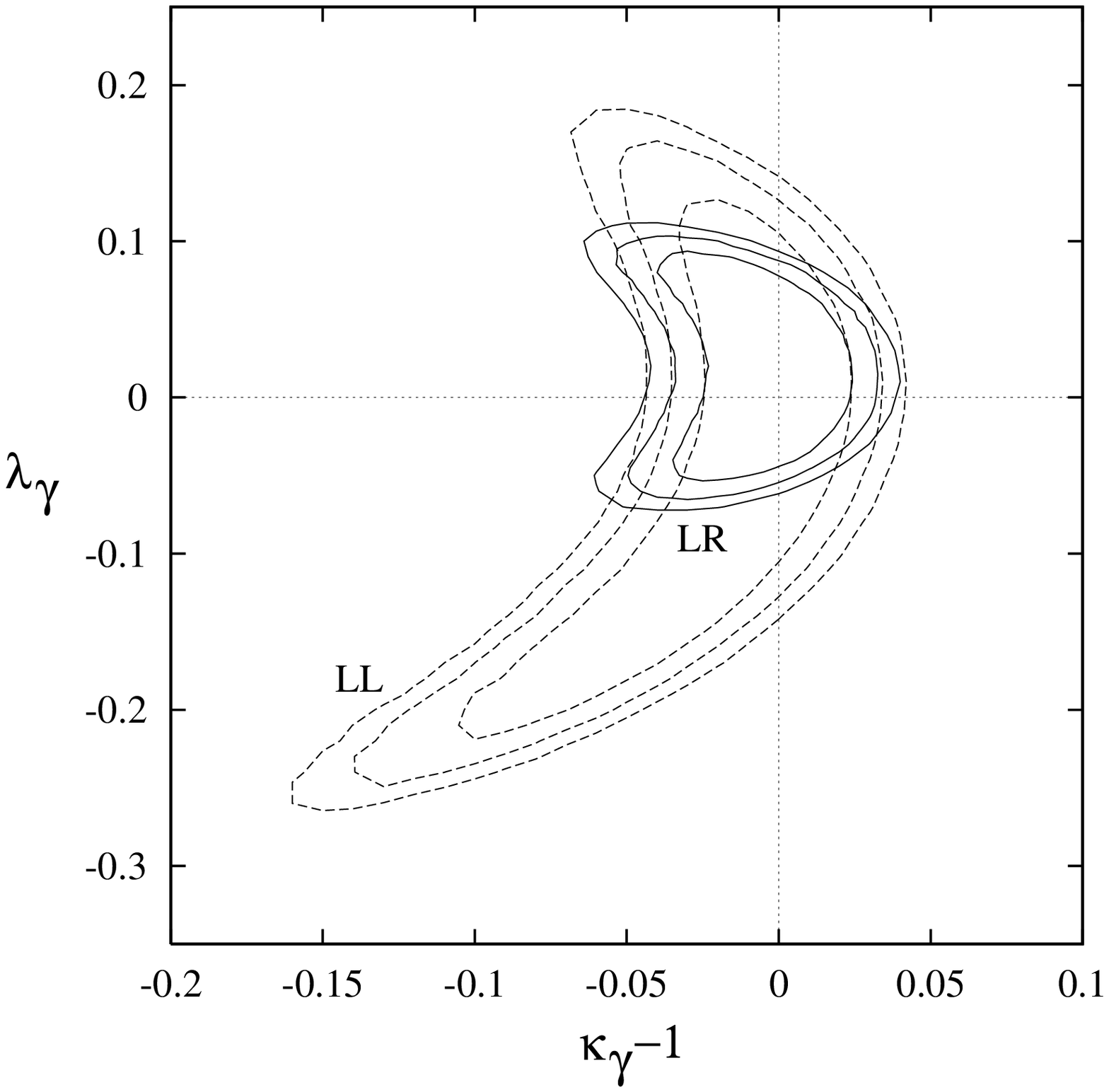}}

\vspace{-30ex}
\caption{As in Fig.~\protect\ref{kgkz}, but in  the $\kg$--$\lg$
         plane instead.}
       \label{kglg}
\end{figure}

\begin{figure}[htb]
\vskip 10in\relax\noindent\hskip -1.2in\relax{\includegraphics{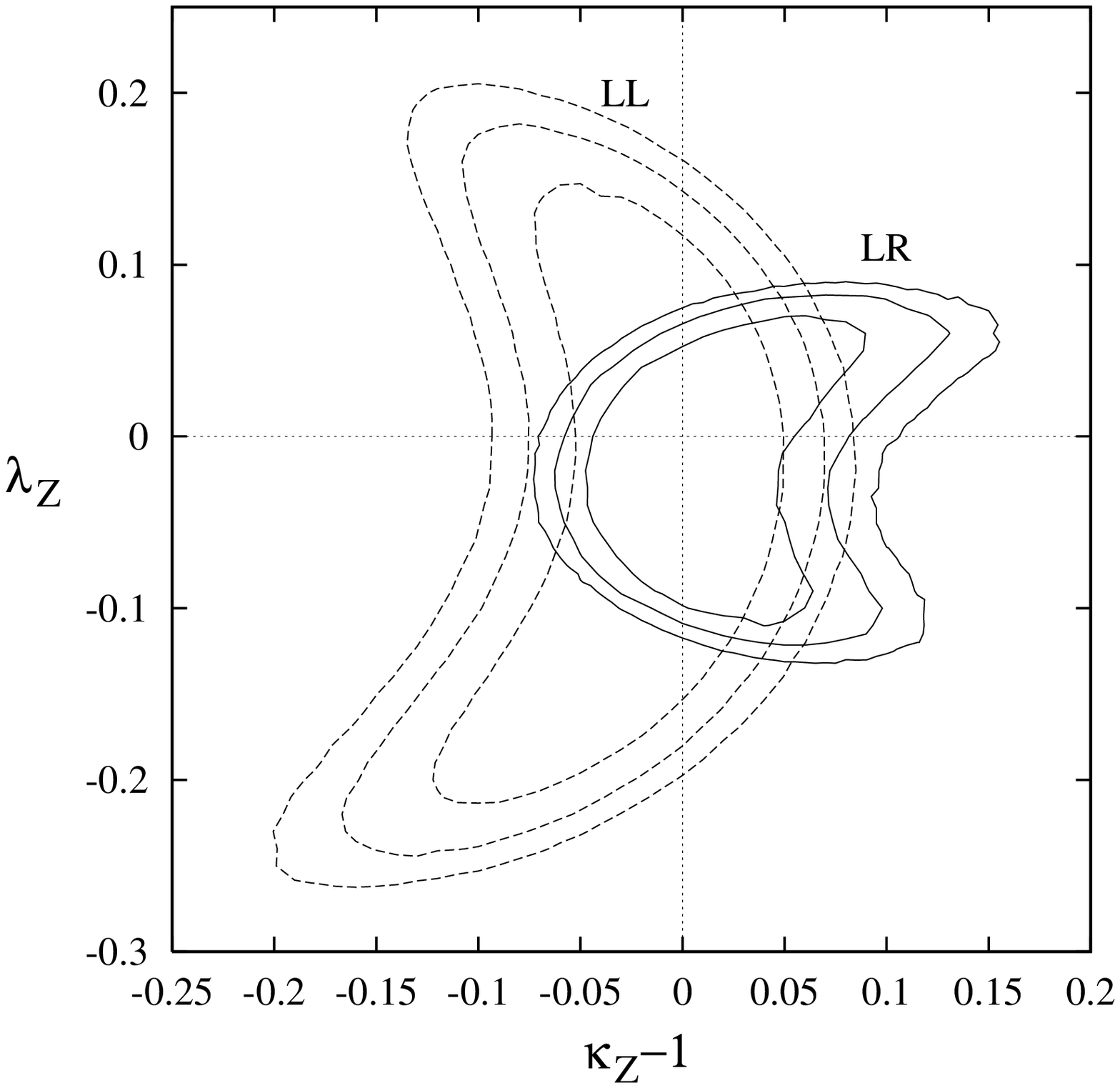}}

\vspace{-30ex}
\caption{As in Fig.~\protect\ref{kgkz}, but in  the $\kZ$--$\lZ$
         plane instead.}
       \label{kzlz}
\end{figure}

\begin{figure}[htb]
\vskip 10in\relax\noindent\hskip -1.2in\relax{\includegraphics{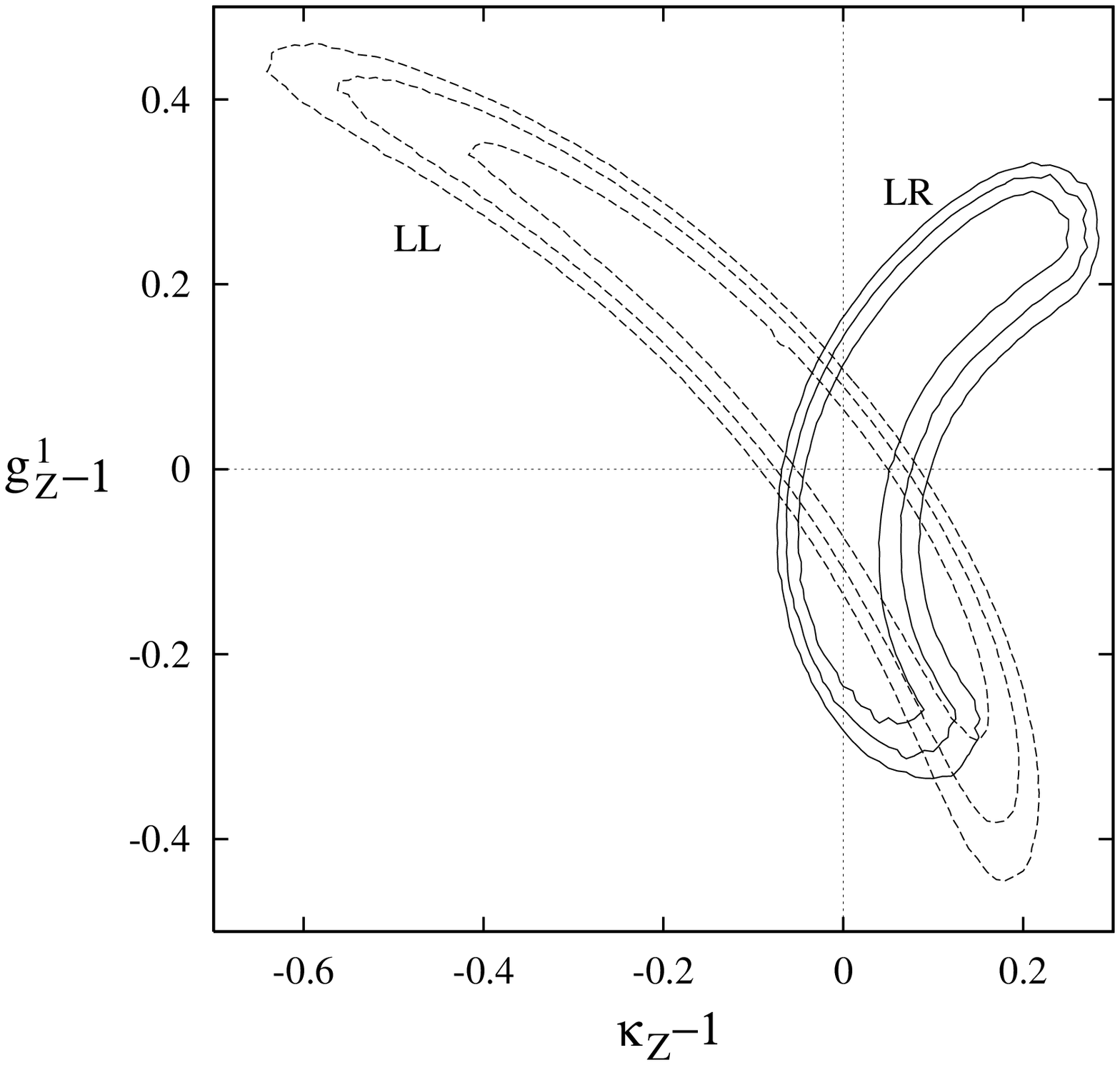}}

\vspace{-30ex}
\caption{As in Fig.~\protect\ref{kgkz}, but in  the $\kZ$--$\gZ$
         plane instead.}
       \label{kzgz}
\end{figure}

We present our results in terms of the $\chi^2$ function (\ref{eq2}),
for different combinations of two anomalous gauge couplings:
$$
(\kg,\kZ)
\qquad
(\lambda_\gamma,\lambda_Z)
\qquad
(\kg,\lambda_\gamma)
\qquad
(\kZ,\lambda_Z)
\qquad
(\kZ,g^1_Z)
$$
For each of these combinations
the other three parameters are kept to their \sm\ value.
The corresponding $\chi^2 = 2,4,6$ contours
are displayed in Figs~\ref{kgkz}-\ref{kzgz}.
Their probabilistic interpretation is the following\footnote{
	It has to be taken with a grain of salt, though,
	since the problem is not linear
	and the errors are not necessarily gaussian.
	This becomes particularly clear in Figs~\ref{kglg}-\ref{kzgz}.
}:
The areas of the parameter space lying outside each contour
can be explored with
63.2\%, 86.5\%\ and 95\%\ confidence respectively.
If only one of the parameters is of interest,
irrespective of the values taken by the other,
the half-planes beyond the edges of the three contours
can be explored with
84.3\%, 95.4\%\ and 98.6\%\ confidence respectively.
If one of the parameters is known to take its \sm\ value,
the latter confidence levels are valid
for the half-lines bounded by the intersections
of the contours with the axis.

Taken individualy,
the $LL$ and $LR$ combinations of initial helicities
show a strong correlation
of the $\kappa_{\gamma,Z}$ and $\lambda_{\gamma,Z}$ parameters\footnote{
	As matter of fact,
	if the electronic angular distribution is not taken into account,
	this correlation assumes a dramatic form
	for the $\kappa$ parameters \cite{wo27}:
	a large closed band of values of the parameters
	cannot be distinguished from the \sm\ case.
}.
Fortunately,
the combined $LL$ and $LR$ data
almost entirely lifts this correlation
and provides much tighter bounds on the parameters.
The same phenomenon also takes place
in the $(\kappa_Z,\lambda_Z)$ and $(\kappa_Z,g^1_Z)$ planes,
to a lesser extent, though.
In spite of its lower overall \xs,
the $LR$ beam polarization
generally yields tighter bounds on the parameters.
This can be traced back to the absence of the $W^-$ exchange diagram,
which otherwise dominates
but is not sensitive to anomalous couplings.
A reduction of the systematical error
would greatly improve the $LL$ results.

A comparison of the resolving power at 90\%\ confidence level
($\chi^2 = 2.71$)
of different experiments
is given in Table~\ref{t1}.
Here we only consider one anomalous coupling at a time,
all others being assumed to take their \sm\ value.
Although this assumption is not realistic,
it allows at least a simple comparison
and may help to provide guidelines\footnote{Note that the strong upper limit
on $\kZ$ obtained in ref.\protect\cite{kane} is not from direct observability,
but rather deduced from radiative correction bounds. Also the authors postulate
relations between the various couplings.}.

It appears that  HERA, LEP2 and TEVATRON are
unlikely to provide competitive results.
In contrast,
the high energies obtained at LHC
make it the best tool for studying higher dimensional operators,
{\em i.e.} the $\lambda$ parameters.
None of the four possible operation modes
of a \lc\ of the next generation
seems to dominate the others.
Note,
however,
that the $e^-\gamma$ and $\gamma\gamma$ modes
do not probe the $Z^0$ sector.
This can be seen as a disadvantage if considered individually,
but certainly constitutes a major advantage
when the information gathered from each experiment
is combined.
Indeed,
it is important to remember that,
if at all,
most likely several or all anomalous gauge parameters
would assume non-trivial values.
It might, therefore,
be dangerous to discard one particular experiment as being insensitive,
only on the basis of the comparison in Table~\ref{t1}.

\section{Conclusion}
    \label{concl}

We have demonstrated
that \ee\ scattering provides a useful reaction
to investigate the gauge sector of the \sm.
Indeed,
slight modifications of the trilinear
$WW\gamma$ and $WWZ$ gauge couplings
would show up as measurable deviations
from the \sm\ expectations
of the outgoing electron's angular distribution
in the process
$e^-e^- \to e^-\nu_eW^-$.
A straightforward $\chi^2$ analysis on data
obtained at a typical \lc\ of the next generation,
would reveal or exclude values of the five $C$ and $P$ conserving
anomalous gauge coupling parameters,
which are competitive with those obtained
in \epm, \ep\ and \pp\ collisions.

In view of the large number of parameters at hand
and their possibly interfering effects,
many experiments would be needed to disentangle
their individual contributions.
In this respect,
\ee\ collisions already offer the bonus
that the two possible beam polarizations
yielding the required signal,
probe complementary regions of the parameter space.

Nevertheless,
even in the presence of an unmistakable departure from the \sm\ prediction
in a given experiment,
it is important to remember
that it is only the conjuction of as many experiments as possible
which might provide the means to understand the origin
of this anomaly.
This is why \lc s,
which can be operated in \epm, \ee, \ep\ and \pp\ modes,
will be priviledged tools for the study of anomalous gauge couplings.

\bigskip
\bigskip

We are very much indebted to Edward Boos and Michael Dubinin
for having provided us with the CompHEP software \cite{CHEP1},
which we have used to generate some of the matrix elements
used in this study.

\newpage
\begin{table}[t]
$$
\begin{array}{|c||c|c||c|c||c|c||c|c||c|c|}
\hline
{\rm machine~or} & \multicolumn{2}{c||}{g^1_Z} &
     \multicolumn{2}{c||}{\kappa_\gamma} & \multicolumn{2}{c||}{\kappa_Z} &
     \multicolumn{2}{c||}{\lambda_\gamma} & \multicolumn{2}{c|}{\lambda_Z}
\\
\cline{2-11}
\rm experiment & \rm min & \rm max & \rm min &
    \rm max & \rm min & \rm max & \rm min & \rm max & \rm min & \rm max
\\ \hline \hline
\rm UA2~\cite{UA2,rahul-1} &  &  & \bf -3.1 & \bf 4.2 &  &  & \bf -3.6
          & \bf 3.5 &  &
\\ \hline
\rm TEVATRON~\cite{rahul-1} &  &  & \bf -2.4 & \bf 3.7 &  &  &  &  &  &
\\ \hline
\rm HERA~\cite{Snowmass} &  &  & 0.5 & 1.5 &  &  & -2 & 2 &  &
\\ \hline
\rm TEVATRON~\cite{kane} &  &  & 0.5 & 1.8 & 0.2 &       &
                                       -0.2 & 0.2 & -0.4 & 0.4
\\ \hline
\rm LHC~\cite{kane} &  &  & 0.8 & 1.2 & 0.8 &     &
                                       -0.02 & 0.02 & -0.03 & 0.03
\\ \hline
\rm LEP2~\cite{kane} &  &  & 0.86 & 1.87 & 0.76 &  & -0.4 & 0.4 &
                            -0.4 & 0.4
\\ \hline
\rm LC500~e^+e^-~\cite{photon} &  &  & 0.985 & 1.14 &  &  & -0.02 & 0.04 &  &
\\ \hline
\rm LC500~e\gamma~\cite{photon} &  &  & 0.96 & 1.04 &  &  & -0.05 & 0.05 &  &
\\ \hline
\rm LC500~\gamma\gamma~\cite{photon} &  &  & 0.98 & 1.015 &  &  & -0.04
   & 0.075 &  &
\\ \hline
\rm LC500~e^-e^- & 0.91 & 1.07 & 0.985 & 1.015 & 0.96 & 1.04 & -0.045 & 0.075
   & -0.11 & 0.06
\\ \hline
\end{array}
$$
\caption{
Parameter values which can be tested by a particular experiment at 90\% C.L.
The boldfaced numbers correspond to limits already set.
For the generic 500 GeV linear collider LC500,
we have assume an integrated luminosity of 10 fb$^{-1}$
for electron or photon beams.
For the $e^-e^-$ option
we have used the combined information from $LL$ and $LR$ beam polarizations.
}
\label{t1}
\end{table}

\newpage

\end{document}